\documentclass[10pt,letterpaper]{article}
\usepackage[top=0.85in,left=1.5in,footskip=0.75in]{geometry}

\usepackage{amsmath,amssymb,amsbsy,amstext}
\usepackage{wasysym}
\usepackage{esint}

\usepackage{changepage}

\usepackage[utf8x]{inputenc}

\usepackage{textcomp,marvosym}

\usepackage{cite}

\usepackage{nameref,hyperref}

\usepackage{microtype}
\DisableLigatures[f]{encoding = *, family = * }

\usepackage[table]{xcolor}

\usepackage{array}

\usepackage{float}

\makeatletter
\renewcommand{\@biblabel}[1]{\quad#1.}
\makeatother

\usepackage{lastpage,fancyhdr,graphicx}
\usepackage{epstopdf}

\begin{document}

\begin{flushleft}
{\Large
\textbf\newline{Tropical Tree Cover in a Heterogeneous Environment: a Reaction-diffusion Model} 
}
\newline
\\
Bert Wuyts\textsuperscript{1,2,3,*},
Alan R. Champneys\textsuperscript{3},
Nicolas Verschueren\textsuperscript{1,3},
Jo I. House\textsuperscript{4},
\\
\bigskip
\textbf{1} College of Engineering, Mathematics and Physical Sciences,
University of Exeter, Exeter 44QF, United Kingdom
\\
\textbf{2} Bristol Centre for Complexity Sciences, University of
Bristol, Bristol BS28BB, United Kingdom
\\
\textbf{3} Applied Nonlinear Mathematics, University of Bristol,
Bristol BS81UB, United Kingdom
\\
\textbf{4} School of Geography, University of Bristol, Bristol BS81SS,
United Kingdom
\\
\bigskip

\noindent \textsuperscript{*}Corresponding author; e-mail: b.wuyts@ex.ac.uk

\bigskip{}

\textit{Manuscript elements:} Figure \ref{fig:Front-between-forest},
Figure \ref{fig:1D-simulation-results.}, Figure \ref{fig:Simulation-results-of},
Figure \ref{fig:Cycles-in-the}, Table \ref{tab:Model-parameters-of-1},
Supporting Information \ref{sec:Model-construction}, and \ref{sec:Forest-growth-rate}
(including Figure \ref{fig:Homogeneous-steady-states}, Figure \ref{fig:Recovery-time-of}
and Table \ref{tab:Model-parameters-of}). 

\bigskip{}


\bigskip{}

\end{flushleft}

\newpage{}

\section*{Update 05/2021: Gradient system requirement}

We describe here briefly the general conditions under which front pinning is expected to occur in the reaction-diffusion system with weakly heterogeneous forcing and show how this applies to the forest-savanna reaction-diffusion model. A reaction-diffusion system will have travelling front solutions if it is a gradient system, i.e. it can be derived from a free energy in the following manner:

\begin{equation}
\partial_t \bold{u}=\bold{f}(\bold{u};p)+\bold{D}\nabla^2\bold{u}=-\boldsymbol{\nabla}_\delta{\cal F}[\bold{u};p],\label{eq:varcond}
\end{equation}
where $\bold{u}$ is an $n$-dimensional vector and $\boldsymbol{\nabla}_\delta=(\delta/\delta u_1,...,\delta/\delta u_n)$ is the functional gradient. For scalar systems ($n=1$), (\ref{eq:varcond}) is always fulfilled. When $n>1$, condition (\ref{eq:varcond}) is usually too strong, as it implies a symmetric Jacobian. However, if (\ref{eq:varcond}) is fulfilled for a subsystem of the reaction-diffusion system, this subsystem is a gradient system, while the whole system is not. 

Writing the forest-savanna model as follows, 
\begin{align*}
\partial_{t}S	&=	r_{s}GT-\omega\left[G+\gamma(S+T)\right]S-m_{S}S-r_{F}SF+D_{S}\nabla^{2}S, \\
\partial_{t}T	&=	\omega\left[G+\gamma(S+T)\right]S-m_{T}T-r_{F}TF, \\
\partial_{t}F	&=	r_{F}(G+S+T)F-b\phi\left[G+\gamma(S+T)\right]F-m_{F}F-cF+D_{F}\nabla^{2}F, 
\end{align*}
where we used the original notation of \cite{Staver2012} for the fire occurrence and savanna tree recruitment functions $\phi$ and $\omega$ and further added their $\gamma$ parameter, which indicates to which extent areas with savanna trees also
conduce fires. We also wrote the parameters in lower case for easier legibility. When substituting $G=1-S-T-F$ and choosing $\gamma=1$, we obtain
\begin{align*}
\partial_{t}S	&=	r_{s}(1-S-T-F)T-\omega\left[1-F\right]S-m_{S}S-r_{F}SF+D_{S}\nabla^{2}S, \\
\partial_{t}T	&=	\omega\left[1-F\right]S-m_{T}T-r_{F}TF, \\
\partial_{t}F	&=	r_{F}(1-F)F-b\phi\left[1-F\right]F-m_{F}F-cF+D_{F}\nabla^{2}F. 
\end{align*} 
We can now see that the forest equation does not depend on any other species than 
forest, making it a scalar reaction-diffusion equation, such that
\[
\partial_{t}F=-\delta {\cal F}\left[F\right]/\delta F,
\]
i.e., there exists a free energy function from which the dynamics of $F$ can be derived. This free energy function is 
\[
{\cal F}[F]=\int_{\Omega}[\frac{D_F}{2}|\nabla F|^{2}+V(F;p)]d^2\mathbf{x},
\]
where $V(F;p)$ is the potential energy function of the forest-grass model derived in this paper (also known as the free energy density in physics) and $p$ is the set of parameters.

Hence, when $\gamma=1$, we can use the Maxwell point of the forest-grass system to predict 
the forest-nonforest front. When $\gamma\neq1$, the Maxwell point will only predict the forest-nonforest front when savanna tree cover near the forest front is sufficiently low,
as described in this paper.

\newpage{}

\section*{Abstract}

Observed bimodal tree cover distributions at particular environmental
conditions and theoretical models indicate that some areas in the
tropics can be in either of the alternative stable vegetation states
forest or savanna. However, when including spatial interaction in
nonspatial differential equation models of a bistable quantity, only
the state with the lowest potential energy remains stable. Our recent
reaction-diffusion model of Amazonian tree cover confirmed this and
was able to reproduce the observed spatial distribution of forest
versus savanna satisfactorily when forced by heterogeneous environmental
and anthropogenic variables, even though bistability was underestimated.
These conclusions were solely based on simulation results for one
set of parameters. Here, we perform an analytical and numerical analysis
of the model. We derive the Maxwell point (MP) of the homogeneous
reaction-diffusion equation without savanna trees as a function of
rainfall and human impact and show that the front between forest and
nonforest settles at this point as long as savanna tree cover near
the front remains sufficiently low. For parameters resulting in higher
savanna tree cover near the front, we also find irregular forest-savanna
cycles and woodland-savanna bistability, which can both explain the
remaining observed bimodality.

\newpage{}

\section*{Introduction\label{sec:Introduction}}

First analyses of the satellite-derived MODIS Vegetation Continuous
Fields (VCF) tree cover product \cite{Townshend2011} found strong
evidence for the bistability hypothesis \cite{Hirota2011,Staver2011b}.
They did this by showing that tropical tree cover data are multimodal
at intermediate rainfall values, i.e. they have multiple maxima in
their empirical probability distribution function. When taking the
plausible assumption that more frequently observed tree cover values
are more stable, such multimodality implies multistability. \cite{Staver2011b}
found forest-savanna bistability, from the observation that the tree
cover data has a bimodal distribution in a rainfall range of intermediate
rainfall, with as modes savanna (about 20\% tree cover) and forest
(about 80\% tree cover). Similarly, \cite{Hirota2011} found forest-savanna-treeless
tristability, with an extra treeless state (about 0\%). The treeless
state was not found by \cite{Staver2011b}, most likely because they
excluded areas with bare soil. A scatterplot of tree cover versus
rainfall revealed how the stability of the states depends on rainfall.
In such a scatterplot, the modes - stable states according to the
dynamical interpretation - show up as regions with high point density.
With increasing mean annual rainfall, the inferred probability of
being in a higher tree cover mode increases. Hence it was concluded
that rainfall can be seen as the bifurcation parameter in a dynamical
system with a hysteresis loop. From here, we restrict our focus to
forest-savanna bistability.

If the bistability model is valid, the low density regions between
the modes indicate instability due to positive feedbacks. To explore
the potential mechanisms driving the positive feedback between savanna
and forest and to check whether there are additional forcing variables,
\cite{Staver2011b} set up a nonlinear statistical model of tree cover
with as predictors mean annual rainfall, dry season length, soil sand
content and fire occurrence. They found that both savanna and forest
can exist in a regime with mild seasonality ($<$7 dry months) and intermediate
rainfall (1000-2500mm/y). In this regime, forest occurrence is highly
predictable from recent fire occurrence, suggesting that fire is an
important factor that can explain the positive feedback between the
savanna and forest states. The hypothesized mechanism in savannas
involves a feedback between grassy cover and fire spread. Fire spread
requires a spatially well-connected grassy fuel layer that occurs
only below a certain tree cover threshold; below this threshold, fire
spread opens up the canopy more, promoting yet better fire spread.
Such a mechanism is consistent with previous theoretical and empirical
research \cite[and the references therein]{Pausas2017}. The existence
of bistability implies that shocks such as forest clearance or drought
could lead to a dramatic increase of fire occurrence and tip an area
of forest into a savanna state. This area of savanna would then remain
locked until large enough increases of rainfall or release of human
pressures allow forests to grow back faster than they are lost by
intermittent fires. 

However, because the empirical studies that support the bistability
hypothesis \cite{Staver2011b,Hirota2011} only rely on spatial data,
bimodality could be a result of confounding factors related to spatial
heterogeneity of climate, plant physiology, soils, human impact, etc.
\cite{Ratajczak2012,Good2016,vanNes2014}. Indeed, in our recent work
\cite{Wuyts2017}, we showed that, at least in the Amazon region,
much of the bimodality is most likely not a consequence of bistability
but of spatial heterogeneity due to factors other than rainfall, including
rainfall seasonality, soils and human impact. Nonetheless, some bimodality
remained in the data, which might still indicate existence of bistability,
albeit on smaller scales than claimed previously. One earlier empirical
study \cite{Staal2016} explored the possibility of more limited bistability
than initially inferred. That they still found wide bistability ranges
is most likely because they only considered the separate instead of
the joint effect of rainfall and seasonality and because they controlled
for fewer confounding factors.

Models of tropical tree cover bistability have remained nonspatial
\cite{Staver2012,vanNes2014,vanNes2018}. However, interaction between
patches is known to be important in tropical forests and savannas,
via processes such as seed dispersal, fire spread and water recycling.
When allowing spatial interaction under the form of diffusion in single-species
reaction-diffusion models with a bistable reaction term, hysteresis
and bimodality disappear; instead, there is an environmentally determined
point that separates both states \cite{Leemput2015,Meron2015,Murray2002}.
Only under the environmental conditions at this point, coined the
Maxwell point (MP), can both states coexist. The MP is a well-understood
concept in\textcolor{black}{{} phase transitions theory} \cite{Huang2009},
used in e.g. materials science, plasma physics and mathematical biology.
In such applications, it is the point of external conditions (e.g.
pressure or temperature) where two separate equilibrium phases of
the considered system have the same free energy. Away from the MP,
there is always one state that has lower free energy. If the system
is spatially homogeneous, perturbations (either diffusion or stochastic
effects) will cause invasion fronts by which the state with the lowest
free energy will perpetuate throughout the domain. When there is a
gradient of external conditions, the front between the stable steady
states pins (i.e. it settles) at the MP \cite{Leemput2015,Wuyts2017}.
This is exactly what we found in our recently developed spatiotemporal
model for Amazonian tree cover \cite{Wuyts2017}, which consists of
a system of equations for several vegetation cover types, including
forest and savanna tree cover. While the model without diffusion produces
bistability between tree cover states, the spatial model did not produce
bistability, but a sharp forest-savanna front, being a function of
mean annual rainfall, rainfall seasonality, soils and human impact.
Taken together with the limited bimodality in the Amazonian data,
this suggests that Amazonian tree cover dynamics can be modeled reasonably
well with a single reaction-diffusion equation exposed to heterogeneous
external conditions. Nonetheless, the limited amount of remaining
bimodality in the data indicates that global bistability, i.e. bistability
despite spatial interaction, may still play a role. Alternatively,
bimodality can also have arisen from endogenously generated cyclic
behavior \cite{Touboul2018,Staver2012}, with cycle periods up to
centuries or millennia, posing a real challenge to falsification of
the model \cite{Touboul2018}, not least because climatic forcing
changes on the same time scales.

Here, we present an analysis of our reaction-diffusion model of tropical
tree cover first used in the simulations of \cite{Wuyts2017}. We
did not include noise terms as noise was treated extensively in \cite{Touboul2018}.
This model is an expansion of the nonspatial bistability model by
\cite{Staver2012} through inclusion of spatial effects (diffusion
and heterogeneity) and human intervention. In this paper, we refer
to the model without savanna trees {[}$S,T=0;F\neq0$ in (\ref{eq:FULL}){]}
as the forest model and to the full model with savanna trees {[}$S,T,F\neq0$
in (\ref{eq:FULL}){]} as the forest-savanna model. We focus in this
work on the analytical derivation of the MP in the homogeneous forest
model and its comparison to the front location in the heterogeneous
forest model and to simulation results of the heterogeneous forest
and forest-savana models. We will show that the MP of the homogeneous
forest model is a good predictor of the front between forest and nonforest
in the heterogeneous forest-savanna model when savanna tree presence
is low. With increasing savanna tree presence, the MP becomes decreasingly
accurate at predicting the front. In this regime, savanna-woodland
bistability and forest-savanna cycles occur, as shown earlier by \cite{Touboul2018}.
We further show that in the spatial model, the savanna-woodland bistability
persists and the forest-savanna cycles can turn irregular. 

\section*{Methods\label{sec:Methods}}

\subsection*{Forest-savanna model}

The full system of partial differential equations representing cover
types as a function of space and time, hereafter referred to as the
forest-savanna model, can be written as
\begin{eqnarray}
\partial_{t}S & = & R_{s}(1-S-T-F)T-Q_{0}[1-h\Phi(T,F)]S-M_{S}S-R_{F}SF+D_{S}\nabla^{2}S,\nonumber \\
\partial_{t}T & = & Q_{0}[1-h\Phi(T,F)]S-M_{T}T-R_{F}TF,\nonumber \\
\partial_{t}F & = & R_{F}(1-F)F-b\Phi(T,F)F-M_{F}F-CF+D_{F}\nabla^{2}F,\label{eq:FULL}
\end{eqnarray}
where 
\begin{eqnarray}
\Phi(T,F) & = & \frac{\tau^{-1}Y_{c}^{4}}{Y_{c}^{4}+(T+F)^{4}},\label{eq:phiYc}
\end{eqnarray}
and $S$ is savanna sapling cover, $T$ savanna adult tree cover,
$F$ forest tree cover, and $\Phi$ fraction of area burnt. This model
can be obtained by starting from the model of \cite{Staver2012} and
adding diffusion terms and human impact. $R_{Y},M_{Y}$ are growth
and mortality rates for $Y\in\{S,T,F\}$. $Y_{c}$ is the critical
value below which fire spread occurs and $\tau$ the maximum fire
return time. $Q_{0}(1-h\Phi)$ is the recruitment rate of savanna
saplings into adult savanna trees; a linearly decreasing function
of burnt area fraction $\Phi$. $b$ is the sensitivity of forest
tree cover to fire, which we choose to be constant here. The forest
removal rate $C$ is a function of distance from human cultivation
$z$, or $C=C(z)$. $\Phi$ is burnt area fraction, which is a monotonic
decreasing and sigmoid-shaped function of nonherbaceous cover $1-G-S=T+F$. 

We show a systematic way for deriving the model (\ref{eq:FULL}) in
Supporting Information \ref{sec:Model-construction}. In our previous
treatment, we included spatial heterogeneity by letting $R_{Y},M_{Y}$
and $Y_{c}$ be functions of natural environmental forcing variables,
such as climate and soils (Table \ref{tab:Model-parameters-of}),
which in turn depend on space. In this work, we strive to make mathematical
analysis as simple as possible, while keeping the model's essential
features. Therefore, we keep rainfall seasonality and soils fixed
at their average values, leading to parameters that are only a function
of mean annual rainfall $P$ or distance to human cultivation $z$.
The resulting simplified functional forms and parameter values are
shown in Table \ref{tab:Model-parameters-of-1}. By assuming that
growth rate saturates to a constant maximum $r_{Y}$ and mortality
stabilizes to a constant minimum $m_{0,Y}$ where water limitation
is less severe, we have chosen
\begin{eqnarray*}
R_{Y}(P) & = & \max[0,r_{Y}(1-e^{-k_{R_{Y}}P+a_{R_{Y}}})],\\
M_{Y}(P) & = & m_{o,Y}+e^{-k_{M_{Y}}P+a_{M_{Y}}},
\end{eqnarray*}
where for $R_{Y}$, $Y\in\{S,F\}$ and for $M_{Y}$, $Y\in\{S,T,F\}$.
$k_{i}$ controls the steepness of the functions and $a_{i}$ the
horizontal position on the $P$ axis. Finally, we took 
\[
Y_{c}(P)=\text{max}[0,Y_{c,0}+k_{c}P],
\]
where $Y_{c,0}>0$ and $k_{c}<0$. $Y_{c}(P)$ captures the assumed
decreasing percolation threshold (critical value of $T+F$) with rainfall.
In drier environments, the effective connectivity between areas in
space is higher, leading to a higher value of tree cover where fire
spread becomes important. 

To introduce spatial heterogeneity, and having already chosen how
$R_{Y},M_{Y}$ and $Y_{c}$ depend on $P$, we still have to choose
how $P$ depends on space. We do this by taking 
\begin{equation}
P(x)=x.\label{eq:Px}
\end{equation}
The resulting rainfall gradient of $1\text{mm/km}$ lies in the range
of what can be expected in the tropics. 

\noindent \begin{flushleft}
\begin{table}[!ht]
\begin{adjustwidth}{-1in}{0in}
\noindent \begin{centering}
\begin{tabular}{cccc}
\hline 
process and equation & parameter & value & units\tabularnewline
\hline 
cover expansion rate & $r_{S},r_{F}$ & 0.09,0.20 & $\text{y}^{-1}$\tabularnewline
$R_{Y}(P)=\max[0,r_{Y}(1-e^{-k_{R_{Y}}P+a_{R_{Y}}})]$ & $k_{R_{S}},k_{R_{F}}$ & 0.005,0.003 & $\text{mm}^{-1}$\tabularnewline
 & $a_{R_{S}},a_{R_{F}}$ & 0.25,1.54 & -\tabularnewline
cover reduction rate by drought & $m_{S,o}=m_{T,o},m_{F,o}$ & 0.023,0.041 & $\text{y}^{-1}$\tabularnewline
$M_{Y}(P)=m_{Y,o}+e^{-k_{M_{Y}}P+a_{M_{Y}}}$ & $a_{M_{S}}=a_{M_{T}},a_{M_{F}}$ & -,-2.15 & -\tabularnewline
 & $k_{M_{S}}=k_{M_{T}},k_{M_{F}}$ & 0.008,0.008 & $\text{mm}^{-1}$\tabularnewline
\begin{tabular}{c}
savanna tree cover recruitment rate\tabularnewline
$Q(\Phi)=Q_{0}(1-h\Phi)$\tabularnewline
\end{tabular} 
& $Q_{0},h$ & 0.04,0.85 & $\text{y}^{-1},\text{-}$\tabularnewline
\begin{tabular}{c}
burnt area fraction\tabularnewline
$\Phi(T,F;P)=\frac{1}{\tau}\frac{Y_{c}^{n}}{Y_{c}^{n}+(T+F)^{n}},$\tabularnewline
\end{tabular} & $\tau,n$ & 2.7,4 & $\text{y},\text{-}$\tabularnewline
critical cover value for fire spread  & $Y_{c,0}$ & 0.56 & -\tabularnewline
$Y_{c}(P)=\text{max}[0,Y_{c,0}+k_{c}P]$ & $k_{c}$ & -1.43e-04 & $\text{mm}^{-1}$\tabularnewline
forest cover fire sensitivity & $b$ & 0.46 & -\tabularnewline
\begin{tabular}{c}
deforestation rate\tabularnewline
$C(z)=ce^{-k_{C}z}$\tabularnewline
\end{tabular} & $c,k_{C}$ & 0.092,0.0015 & -,$\text{m}^{-1}$\tabularnewline
diffusion coefficient of $S,F$ & $D_{S},D_{F}$ & 0.2,0.1 & $\text{km}^{2}\text{y}^{-1}$\tabularnewline
\hline 
\end{tabular}
\par\end{centering}
\caption[Model parameters]{Model parameters and functional forms of the forest-savanna model
when fixing rainfall seasonality and soils at their average (\ref{eq:FULL}).
These were obtained by filling in the average for rainfall seasonality
and soils in the equations of Table \ref{tab:Model-parameters-of}.}
\label{tab:Model-parameters-of-1}
\end{adjustwidth}
\end{table}
\par\end{flushleft}

\subsection*{Forest model (S,T=0)\label{subsec:Forest-grass-model-(B,S,T=00003D0)}}

We now set up the spatial model of forest cover (and its complement
$1-T$, grass cover). This is done by setting $S=T=0$ in (\ref{eq:FULL}),
leading to
\begin{equation}
\partial_{t}F=R_{F}(P)F(1-F)-M_{F}(P)F-bF\Phi(F)-C(z)F+D_{F}\nabla^{2}F.\label{eq:FG}
\end{equation}
It will be helpful in the analysis that follows to produce a nondimensional
version of this model. We first take $u=F$ and rescale $t\rightarrow bt/\tau$.
We take as nondimensional constants (see Table \ref{tab:Model-parameters-of-1}),
\[
\rho=\frac{r_{F}}{b}\tau,\mu=\frac{e^{a_{M_{Y}}}}{b}\tau,\mu_{0}=\frac{m_{F,o}}{b}\tau,\gamma=\frac{c}{b}\tau,\delta_{F}=\frac{D_{F}}{b}\tau,
\]
and replace $\kappa_{r}=k_{R_{F}},\kappa_{m}=k_{M_{F}},a=a_{R_{Y}},u_{c,0}=Y_{c,0}$
for lighter notation. We further take as nondimensional functions
\begin{eqnarray*}
r(P) & = & 1-e^{-\kappa_{r}P+a},\\
m(P) & = & e^{-\kappa_{m}P},\\
f(u;P) & = & \frac{u_{c}(P)^{4}}{u_{c}(P)^{4}+u{}^{4}},\\
u_{c}(P) & = & \max[0,u_{c,0}+k_{c}P],\\
c(z) & = & e^{-k_{C}z}.
\end{eqnarray*}
When putting everything together, the following dimensionless form
of the PDE is obtained,
\[
\partial_{t}u=\rho r(P)(1-u)u-\mu m(P)u-uf(u;P)-\gamma c(z)u-\mu_{0}u+\delta_{F}\nabla^{2}u.
\]
With rescaling $x\rightarrow\sqrt{\delta_{F}}x$ we then obtain 
\[
\partial_{t}u=\rho r(P)(1-u)u-\mu m(P)u-uf(u;P)-\gamma c(z)u-\mu_{0}u+\nabla^{2}u.
\]
When making the further substitutions,
\begin{eqnarray*}
\alpha(P,z) & = & \rho r(P)-\mu m(P)-\gamma c(z)-\mu_{0},\\
\beta(P) & = & \rho r(P),
\end{eqnarray*}
we obtain
\begin{equation}
\partial_{t}u=\alpha(P,z)u-\beta(P)u^{2}-uf(u;P)+\nabla^{2}u.\label{eq:FG_diffST}
\end{equation}
We will show that the front between forest and grassland as a function
of the forcing variables can be found analytically. While this model
does not include savanna tree cover, we can compare the forest-savanna
model with this one to see how the presence of savanna trees affects
the results. 

\subsection*{Parameters, simulation and figures \label{subsec:Simulation-set-up-and}}

All parameter values have roughly the same values as those in \cite{Wuyts2017}.
Table \ref{tab:Model-parameters-of} summarizes the parameters and
functions used in the model. The forest growth rate can be easily
inferred from the data (see Supporting Information \ref{sec:Forest-growth-rate}).
We ran the 1D model in MATLAB \cite{MathWorks2012} with the $\texttt{ode45}$
algorithm based on Runge-Kutta 4th and 5th order temporal discretization
(variable $\Delta$t) and central difference spatial discretization
($\Delta x=0.67$), no-flux boundary conditions and random initial
conditions. The chosen left and right boundaries are $0\text{km}$
and $3000\text{km}$.

We made two types of figures: a phase plot with the front location
in parameter space (Figure \ref{fig:Front-between-forest}), and,
scatterplots of cover types versus rainfall in the heterogeneous models
(Figures \ref{fig:1D-simulation-results.} and \ref{fig:Simulation-results-of}).
To create the phase plot of the heterogeneous models in Figure \ref{fig:Front-between-forest},
we needed to extract the rainfall value at the front from the model
output. For the simulations (markers in Figure \ref{fig:Front-between-forest}),
we did this via a robust curve fitting method, fitting the logistic
function (goodness of fit $R^{2}>.9$),
\begin{equation}
F^{*}(P)=\frac{A}{1+\exp[-k(P-P_{f})]},\label{eq:logitfit}
\end{equation}
and extracting $P_{f}$, with the MATLAB \cite{MathWorks2012} curve
fitting tool. In the numerical continuation of the heterogeneous forest
model (solid blue line in Figure \ref{fig:Front-between-forest}),
we did this via 
\begin{equation}
P_{f}=\arg\max F_{x}^{*}(P),\label{eq:argmaxFstar}
\end{equation}
where $F_{x}^{*}$ is the spatial derivative of the front solution.
We used (\ref{eq:logitfit}) instead of (\ref{eq:argmaxFstar}) in
the simulations because in the forest-savanna model, savanna species
can induce gradients of $F$ away from the front. The two methods
give the same results when there are no savanna trees (compare $+$
and solid blue line in Figure \ref{fig:Front-between-forest}).

The analysis of the homogeneous model almost exclusively required
symbolic analysis, which we did with Mathematica \cite{WolframResearchInc2018}.

\section*{Results\label{sec:Results}}

In the first section below, we derive the MP of the homogeneous forest
model. In the second section, the front pinning location in the heterogeneous
forest model is derived via a numerical continuation. The third section
shows simulation results of the heterogeneous forest and forest-savanna
models.

\subsection*{Maxwell point of the homogeneous forest model\label{subsec:Maxwell-point-in}}

For simplicity we shall consider one spatial dimension, which gives
rise to scalar fronts rather than domain boundaries in the form of
line fronts. While the approach in 2D is identical once one chooses
a direction of propagation of any invasion front, front dynamics will,
unlike in 1D, be influenced by front curvature, but this is minimal
for the spatial scales considered \cite{Wuyts2017c,Wuyts2017}. Because
we first assume forcing to be homogeneous, we can further treat $p$
and $z$ as parameters. We further also assume that the front width
is very small compared to the domain size, such that it is justified
to take the domain size as approximately infinite. 

When starting from (\ref{eq:FG_diffST}), hiding the dependence on
$p$ and $z$, grouping common factors, and indicating further $\partial_{t}u$
by $u_{t}$ and $\partial^{2}u/\partial x^{2}$ by $u_{xx}$, we obtain
\begin{equation}
u_{t}=[\alpha-\beta u-f(u)]u+u_{xx}.\label{eq:RDE1D}
\end{equation}
As the nonlinear term causes bistability, we expect traveling front
solutions {[}see e.g. \cite{Pismen2006,Murray2002,Meron2015}{]} between
the stable steady states of the form $u(\xi)$ with $\xi=x-ct$ and
$c$ the wave speed, with boundary conditions $u(-\infty)=u_{-}$
and $u(\infty)=u_{+}$ such that we can rewrite our equation as 
\[
-cu'=[\alpha-\beta u-f(u)]u+u'',
\]
where $u'=du/d\xi$ and $u''=d^{2}u/d\xi^{2}$. When multiplying by
$u'$, we obtain
\[
-c(u')^{2}=[\alpha-\beta u-f(u)]uu'+u''u'.
\]
Integrating this with respect to $\xi$ over the real axis, we further
obtain 
\begin{eqnarray*}
-c\int_{-\infty}^{\infty}(u')^{2}d\xi & = & \int_{-\infty}^{\infty}[\alpha-\beta u-f(u)]uu'd\xi+\int_{-\infty}^{\infty}u''u'd\xi,\\
 & = & \int_{u_{-}}^{u_{+}}[\alpha-\beta u-f(u)]udu-\int_{u_{-}}^{u_{+}}u'du',\\
 & = & \int_{u_{-}}^{u_{+}}[\alpha-\beta u-f(u)]udu-[\frac{1}{2}u'^{2}]_{u_{-}}^{u_{+}}.
\end{eqnarray*}
As the solution is flat at the boundaries, we have $[\frac{1}{2}u'^{2}]_{u_{-}}^{u_{+}}=0$,
such that 
\[
-c\int_{-\infty}^{\infty}(u')^{2}d\xi=\int_{u_{-}}^{u_{+}}[\alpha-\beta u-f(u)]udu.
\]
As the integrand of the left hand side of this expression is always
positive, we have 
\begin{equation}
\text{sign}(c)=\text{-sign}\{\int_{u_{-}}^{u_{+}}[\alpha-\beta u-f(u)]udu\}=\text{sign}(\Delta V),\label{eq:signdV}
\end{equation}
where we have defined, 
\begin{eqnarray*}
\Delta V & \equiv & -\int_{u_{-}}^{u_{+}}[\alpha-\beta u-f(u)]udu\\
 & = & [-\alpha u^{2}/2+\beta u^{3}/3]_{u_{-}}^{u_{+}}+\int_{u_{-}}^{u_{+}}f(u)udu.
\end{eqnarray*}
Hence, we see that the dynamics can be derived from the potential
by 
\[
u_{t}=-V_{u}+\nabla^{2}u.
\]

At the MP, the front is stationary, i.e. $c=0$, such that according
to (\ref{eq:signdV}),
\[
\Delta V=[-\alpha u^{2}/2+\beta u^{3}/3]_{u_{-}}^{u_{+}}+\int_{u_{-}}^{u_{+}}f(u)udu=0.
\]
This allows calculation of an expression for the MP as a function
of the parameters $\alpha$ and $\beta$. These parameters, in turn,
are a function of the external forcings of the model. 

\begin{figure}[!h]
\begin{adjustwidth}{-1.0in}{0in}
\begin{centering}
\includegraphics[width=.75\columnwidth]{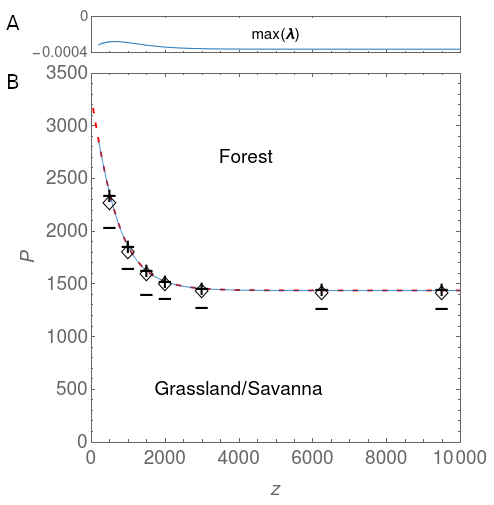}
\par\end{centering}
\caption[Front location and stability.]{Maxwell point in the homogeneous forest model and pinning rainfall in the heterogeneous models. 
(A) Maximum eigenvalue of (\ref{eq:opdiscr}) for a range of $z$
values in the heterogeneous forest model (i.e. along the solid blue
line in panel B). (B) Front between forest and savanna/grassland in
$(P,z)$ space for different models. The dashed red line shows the
theoretically derived MP from the homogeneous forest model and the
solid blue line the location of the forest front in the heterogeneous
forest model obtained by a numerical continuation. Markers show at
which rainfall value the front settles in the heterogeneous models
for given $z$ values: ($+$) forest model (\ref{eq:FG}), ($\diamondsuit$)
forest-savanna model (\ref{eq:FULL}) with $r_{S}=0.09$ and $Q_{0}=0.04$,
($-$) forest-savanna model (\ref{eq:FULL}) with $r_{S}=0.13$ and
$Q_{0}=0.09$. }
\label{fig:Front-between-forest}
\end{adjustwidth}
\end{figure}

If we choose $f(u)$ as in (\ref{eq:BA}) with $Y=u$, 
\begin{equation}
f(u)=\frac{u_{c}^{4}}{u_{c}^{4}+u{}^{4}},\label{eq:BA4}
\end{equation}
then $\int f(u)du$ can be calculated analytically as 
\[
\int f(u)du=\frac{u_{c}^{2}}{2}\text{arctan}[\left(u/u_{c}\right)^{2}],
\]
such that 
\[
V(u)=\frac{\beta u^{3}}{3}-\frac{\alpha u^{2}}{2}+\frac{u_{c}^{2}}{2}\text{arctan}[\left(u/u_{c}\right)^{2}].
\]
$\Delta V=V(u_{+})-V(u_{-})$ can be found analytically if $u_{+}$
and $u_{-}$ can be found analytically. However, $u_{+},u_{-}$ can
only be found analytically when the (integer) exponent in (\ref{eq:BA})
is $1\leq n\leq3.$ As we chose $n=4$, this step has to be done numerically.
From here, the MP can be calculated by finding the root of $\Delta V$
as a function of its parameter(s). Also this is only possible numerically.
The result of this calculation is shown as the dashed red line in
Figure \ref{fig:Front-between-forest}. For the parameters shown in
Table \ref{tab:Model-parameters-of-1}, without human impact, and,
at average rainfall seasonality and soils, the MP of the forest model
lies at a mean annual rainfall of 1438mm. Areas receiving $P>1438\text{mm}$
will experience an invasion of forest while areas receiving $P<1438\text{mm}$
will experience loss of forest. When including human impact, forest
is only considerably affected by human impact when it is less than
$z\approx2\text{km}$ away from agricultural areas. 

Without spatial interaction in the forest model, there is a wide range
where forest is bistable with grassland ($\sim$1200-3500mm, upper
branch and lower zero branch indicated with solid lines in Figure
\ref{fig:Homogeneous-steady-states}). Hence, we showed here that
including spatial interaction causes the bistability range to collapse
into one point - the MP. Note that when there are $N$ forcing variables,
the MP is not a point but a $N-1$ dimensional surface in phase space.
Away from the MP, the only stable state is the one with lowest potential
energy $V$. The alternative state with lower potential energy is
now metastable. It can persist when: (1) the whole spatial domain
is homogeneously in that state, and (2) that this homogeneous state
is not sufficiently perturbed. Nonetheless, neither of these conditions
are easily met in reality.

\subsection*{Front pinning in the heterogeneous forest model}

When external conditions are heterogeneous, the parameters $p,z$
and the solutions $u_{+},u_{-}$ are functions of $x$ and the approach
in the previous section cannot be used any more. However, one can
expect that when the spatial dependence is weak, it can still be used
as an approximation. It can then be expected that in the limit of
$t\rightarrow\infty,$ areas receiving $P>P_{MP}$ will have forest
while areas receiving $P<P_{MP}$ will not have forest, with the front
pinned at $P_{MP}$. The stability of the pinned front solution can
be verified via a linear stability analysis. When writing the reaction
term of (\ref{eq:RDE1D}) as ${\cal R}[u;P]$, we have 
\begin{equation}
u_{t}={\cal R}[u(x);P(x)]+u_{xx}.\label{eq:RDER}
\end{equation}
At the front solution $u=u^{*}(x)$, we perturb the solution with
$\delta u(x,t)\ll1$ and see how this perturbation grows by substituting
$u^{*}(x)+\delta u(x,t)$ and neglecting higher order terms in $\delta u$,
\begin{eqnarray*}
[u^{*}(x)+\delta u(x,t)]_{t} & = & {\cal R}[u^{*}(x)+\delta u(x,t);P(x)]+[u^{*}(x)+\delta u(x,t)]_{xx},\\{}
[\delta u(x,t)]_{t} & = & {\cal R}[u^{*}(x);P(x)]+\frac{\partial{\cal R}}{\partial u}[u^{*}(x),P(x)][\delta u(x,t)]+[u^{*}(x)+\delta u(x,t)]_{xx},\\{}
[\delta u(x,t)]_{t} & = & \frac{\partial{\cal R}}{\partial u}[u^{*}(x),P(x)][\delta u(x,t)]+[\delta u(x,t)]_{xx},
\end{eqnarray*}
where the second step is possible because ${\cal R}(u^{*};P)+u_{xx}^{*}=0$
as $u^{*}$is a solution of (\ref{eq:RDER}). Therefore, the front
solution is only stable with respect to perturbation when all eigenvalues
of the operator,
\begin{equation}
{\cal L}(x)=\frac{\partial R}{\partial u}[u^{*}(x),P(x)]+\partial_{xx},\label{eq:opcont}
\end{equation}
have negative real parts. In our case, it is not possible to obtain
the front solution $u^{*}(x)$ analytically. Therefore, linear stability
can be evaluated numerically, by calculating the eigenvalues of the
discretized form of (\ref{eq:opcont}), which is the $n\times n$
matrix
\begin{equation}
\boldsymbol{{\cal L}}=\frac{\partial{\cal R}}{\partial u}(\mathbf{u}^{*},\mathbf{P})\mathbf{I}+\frac{\mathbf{L}}{\Delta x^{2}},\label{eq:opdiscr}
\end{equation}
where $\mathbf{u}^{*}=[u(x_{0}),u(x_{1}),...,u(x_{n-1})]$ and $\mathbf{P}=[P(x_{0}),P(x_{1}),...,P(x_{n-1})]$
are the discretized front solution and rainfall values as a function
of space, with $x_{k}=x_{0}+k\Delta x$, $\mathbf{L}/\Delta x^{2}$
the discretized Laplacian, and $\mathbf{I}$ the identity matrix.
If we define $\max\mathbf{v}$ as the maximum of a vector $\mathbf{v}$'s
elements and $\boldsymbol{\lambda}$ as the vector with $n$ eigenvalues
of (\ref{eq:opdiscr}), the condition for stability is hence
\begin{equation}
\max\Re(\boldsymbol{\lambda})<0,\label{eq:maxRelamb}
\end{equation}
where $\Re$ indicates that we take the real part. Because the front
solution $\mathbf{u}^{*}$ depends on all the parameters, $\boldsymbol{{\cal L}}$
is calculated for only one point in parameter space. To obtain information
on the stability of all front solutions in a given parameter range,
one needs to obtain the solution for a set of points in that range
and evaluate $\boldsymbol{{\cal L}}$ for each of them. Starting from
a known front solution, pseudo-arclength continuation \cite{Avitabile2016,Rankin2013,Beyn2000}
allowed us to find other front solutions of (\ref{eq:RDER}) in parameter
space. To compare the results with those of the previous section,
we plot the rainfall value at which the front pins in the heterogeneous
equation as a function of $z$ (distance from human cultivation).
We extracted the location of the front via (\ref{eq:argmaxFstar})
for each value of $z$. Our analysis shows that the front solution
of the heterogeneous forest model (solid blue line in Figure \ref{fig:Front-between-forest})
is indistinguishable from the MP of the homogeneous forest model (dashed
red line in Figure \ref{fig:Front-between-forest}). Moreover, we
found that all for each value of $z$ considered (\ref{eq:maxRelamb})
is satisfied (solid red line in Figure \ref{fig:Front-between-forest}),
indicating that each front solution is a stable steady state, or more
specifically, a stable node, as all eigenvalues of (\ref{eq:opdiscr})
are real. It can hence be concluded that, at least for our setup with
weak spatial dependence, the front of the heterogeneous forest equation
pins at the MP of the homogeneous forest equation.

\subsection*{Simulation of the heterogeneous models}

\begin{figure}[!h]
\begin{adjustwidth}{-1.0in}{0in}
\begin{centering}
\includegraphics[width=1.25\columnwidth]{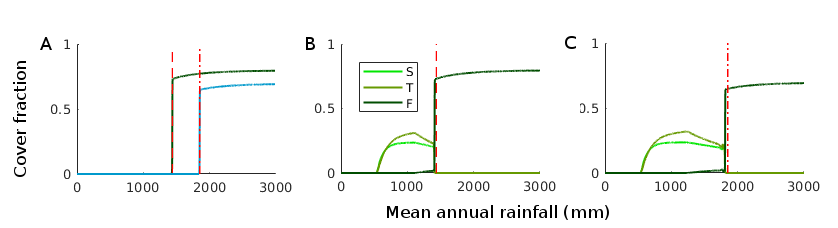}
\par\end{centering}
\caption[1D simulation results with effect human impact for several models.]{Simulation results and the effect of human impact for the models
under low impact of savanna trees ($r_{S}=0.09$,$Q_{0}=0.04$, $\tau=2.7$).
(A) Forest model (\ref{eq:FG}) under natural (green) and impacted
conditions (blue, 1km from cultivated areas). (B) Forest-savanna model
(\ref{eq:FULL}) under natural conditions. (C) Forest-savanna model
(\ref{eq:FULL}) with human impact (1km from cultivated areas). The
red dashed line shows the derived value of the MP in the natural forest
model. The red dash-dotted line shows the derived value of the MP
in the forest model with human impact. Rainfall can also be seen as
a spatial coordinate because the model was forced by heterogeneous
rainfall $P(x)=x$. }
\label{fig:1D-simulation-results.}
\end{adjustwidth}
\end{figure}

\begin{figure}[!h]
\begin{adjustwidth}{-1.0in}{0in}
\begin{centering}
\includegraphics[width=1.25\columnwidth]{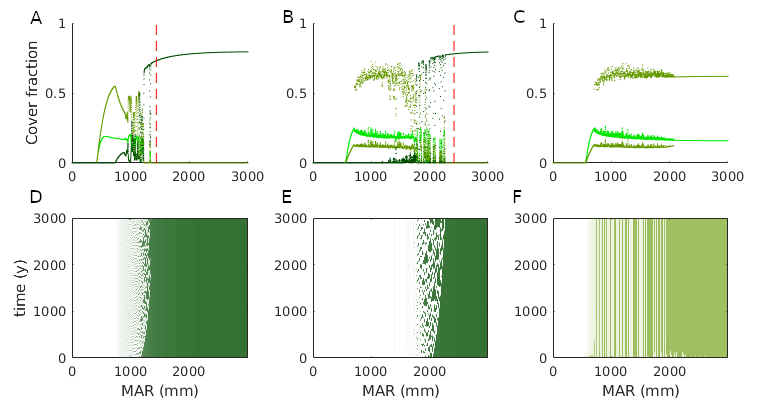}
\par\end{centering}
\caption{Simulation results of the forest-savanna model (\ref{eq:FULL}) with
higher savanna sapling growth rate ($r_{S}=0.13$) and: (A,D) higher
sapling recruitment into adults ($Q_{0}=0.09$), (B,E) higher recruitment
into adults and lower fire return interval ($Q_{0}=0.2$, $\tau=1$).
(C,E) Same as in (B,E) but without forest trees. The upper panels
show cover fraction versus rainfall at the end of the simulation and
of all cover types. The lower panels show forest (D,E) or of savanna
adult tree cover over the spatial domain (with the location indicated
by its rainfall) as a function of time. The MP of the corresponding
forest model is shown with the dashed red line. See Figure 1 for legend.}
\label{fig:Simulation-results-of}
\end{adjustwidth}
\end{figure}

Here we show steady state profiles of vegetation by the heterogeneously
forced models. We remind the reader that the used forcing is a linear
relation between distance from the origin and rainfall (\ref{eq:Px})
such that at the chosen left and right boundaries $P(0\text{km})=0\text{mm}$
and $P(3000\text{km})=3000\text{mm}$, respectively. Therefore, the
x-axis of the plots in Figure \ref{fig:1D-simulation-results.} and
\ref{fig:Simulation-results-of} is both distance from the origin
in $\text{km}$ or mean annual rainfall in mm.

Figure \ref{fig:1D-simulation-results.} shows the steady states of
$F$ as a function of rainfall by the forest model (\ref{eq:FG_diffST})
and of $S,T$ and $F$ by the forest-savanna model (\ref{eq:FULL})
for parameters leading to low savanna tree presence ($r_{s}=0.09$,
$Q_{0}=0.04$, $\tau=2.7$), with and without human impact. Without
human impact, all models have their forest front pinned at a rainfall
value of about 1400mm {[}Figure \ref{fig:1D-simulation-results.}A
(green), B (green){]}, with forest occurring above and grassland or
savanna below this value. Adding human impact results in a shift of
the forest front to higher rainfall values (\ref{fig:1D-simulation-results.}A
blue versus green; \ref{fig:1D-simulation-results.}C vs \ref{fig:1D-simulation-results.}B).
In the forest model, the MP obtained from the analysis of the homogeneous
model (Figure \ref{fig:1D-simulation-results.}, dashed lines without
human impact and dash-dotted line with human impact) accurately predicts
the location of the forest front (Figure \ref{fig:1D-simulation-results.}A).
The model with savanna trees (forest-savanna model) has its forest
front at slightly lower rainfall values than the model without savanna
trees (Figure \ref{fig:1D-simulation-results.}C vs Figure \ref{fig:1D-simulation-results.}B).
The rainfall value at which the front pins is indicated by markers
in Figure \ref{fig:Front-between-forest} for a wider range of $z$
values, confirming the good match {[}perfect match for the forest
model ($+$) and small bias for the forest-savanna model ($\diamondsuit$){]}
between the rainfall value at which the front pins and the MP of the
homogeneous forest model for the parameters chosen here. 

Figure \ref{fig:Simulation-results-of}A shows the cover types versus
rainfall when we choose parameters leading to higher savanna tree
cover ($r_{s}=0.13$, $Q_{0}=0.09$). As before, there is forest on
the wet side and savanna on the dry side of the x axis. However, now
adult savanna trees reach higher cover values and there is a larger
difference between the MP and the location of the front (see $-$
markers in Figure \ref{fig:Front-between-forest} for a wider range
of $z$ values). The MP becomes decreasingly accurate as predictor
of the forest front with increasing savanna tree cover (Figure \ref{fig:Simulation-results-of}B
versus \ref{fig:Simulation-results-of}A and \ref{fig:1D-simulation-results.}D).
Moreover, beyond the point where savanna cover decreases, there is
a range of rainfall values below the forest front where forest and
savanna tree cover show high variation due to irregular oscillations
of forest and savanna tree cover (Figure \ref{fig:Simulation-results-of}D).
Figure \ref{fig:Simulation-results-of}B shows that when savanna tree
recruitment is increased further ($Q_{0}=0.2$) and when also the
fire return interval is decreased ($\tau=1)$, savanna tree cover
becomes bistable below a rainfall of about 1000mm and the range of
rainfall with forest-savanna cycles widens. We will further refer
to the low savanna tree cover state as the savanna state and to the
high savanna tree cover state the woodland state. Note that the savanna
tree cover bistability also occurs (for the same parameters) without
forest trees (Figure \ref{fig:Simulation-results-of}C,F), but up
to a rainfall of about 2500mm. 

\begin{figure}[!h]
\begin{adjustwidth}{-1.0in}{0in}
\begin{centering}
\includegraphics[width=1.\columnwidth]{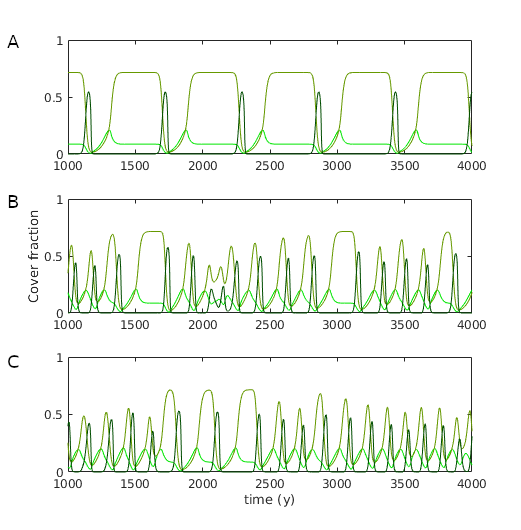}
\par\end{centering}
\caption{Cycles in the forest-savanna model with high savanna tree presence
($r_{S}=0.13$, $Q_{0}=0.2$ and $\tau=1$):
(A) nonspatial model with $P=1500\text{mm}$, (B) spatially homogeneous
model with $P=1500\text{mm}$, (C) Spatially heterogeneous model at
the point where $P=1500\text{mm}$. See Figure \ref{fig:1D-simulation-results.}
for legend. }
\label{fig:Cycles-in-the}
\end{adjustwidth}
\end{figure}

In Figure \ref{fig:Cycles-in-the}, we show the forest-savanna cycles
in more detail. During the cycles, forest tree cover lags behind savanna
tree cover. The changes between states occur over decades, but the
periods of stability between the transitions can persist for several
centuries (or longer, depending on the parameters). The nonspatial
system only produces a regular cycle (Figure \ref{fig:Cycles-in-the}A)
while the spatially homogeneous system with diffusion (Figure \ref{fig:Cycles-in-the}B)
has irregular cycles. The spatially heterogeneous system has similar
irregular cycles (Figure \ref{fig:Cycles-in-the}C). The irregularity
of these cycles can hence be induced by diffusion alone. 

\section*{Discussion\label{sec:Discussion} }

In this paper, we have provided a first analytical and numerical analysis
of our spatially heterogeneous reaction-diffusion model of tropical
tree cover. We have treated this model before with a more realistic
set-up \cite{Wuyts2017} (in 2D, with noise and forced by observed
climate, soil and human impact), but we formulated it here in an as
simple as possible form (in 1D, deterministic and forced by linear
rainfall) for easier mathematical analysis. The heterogeneity was
captured with the relation (\ref{eq:Px}), such that low x values
represent dry and high x values represent wet areas. From the homogeneous
system without savanna trees/saplings {[}$S=T=0$, (\ref{eq:FG_diffST}){]},
a Maxwell point was derived. We showed via a numerical continuation
and linear stability analysis of the spatially heterogeneous forest
model that this MP is still of use for the spatially heterogeneous
case because here, it is the parameter value at which the forest front
pins. The MP of the homogeneous forest model and the rainfall value
at which the forest model's front pins as a function of external parameters
(the dashed red line and the solid blue line in Figure \ref{fig:Front-between-forest}
respectively) are indistinguishable and have the same shape as what
was obtained in \cite{Wuyts2017} by simulation. Existence of a MP
in reaction-diffusion equation with a bistable reaction term \cite{Murray2002,Pismen2006,Meron2015}
and pinning under heterogeneity \cite{Leemput2015} is consistent
with previous work. For parameters that lead to low cover of savanna
trees, the MP of (\ref{eq:FG_diffST}) is also a good predictor of
the forest-savanna model's forest front {[}$S,T\neq0$, (\ref{eq:FULL}){]}
(Figure \ref{fig:1D-simulation-results.}C-F). This is because the
effect of savanna trees on forest trees, mediated by burnt area {[}see
(\ref{eq:FULL}){]}, remains negligible when savanna tree cover near
the forest front stays below the threshold where fire spread is inhibited,
i.e. $T<Y_{c}$. Choosing parameters such that savanna tree cover
near the forest front exceeds this threshold ($T\apprge Y_{c}$) makes
the forest front shift away from the MP of (\ref{eq:FG_diffST}),
towards drier areas (Figure \ref{fig:Simulation-results-of}A,B).
In this regime where savanna tree cover affects forest tree cover,
we also found forest-savanna cycles and savanna-woodland bistability,
which both can lead to bimodal tree cover distributions under the
same external forcings. These cycles are consistent with the existence
of Hopf bifurcations in the nonspatial system \cite{Touboul2018}
above a certain value of the parameters equivalent to $P$ and $r_{s}$.
For an explanation of the physical mechanism behind the cycles, we
refer to \cite{Touboul2018}. We found that the cycles can turn irregular
by diffusion. That the irregular cycles are produced endogenously
suggests that close to the forest front, sudden and unpredictable
loss of forest can occur without climatic or anthropogenic perturbations.
We speculate that the irregularity is due to spatiotemporal chaos,
which is known occur in the wake of traveling fronts \cite{Sherratt1995,Sherratt2008}.
To prove this, it would need to be shown additionally that the cycles
produced by the deterministic system are truly aperiodic and that
there is sensitivity to initial conditions \cite{Strogatz2014}. We
further showed via simulation that bistability of a savanna and a
woodland state can arise in the savanna model (i.e. the model without
forest trees) under a regime of high sapling recruitment and high
fire occurrence (Figure \ref{fig:Simulation-results-of}C,F). When
introducing forest trees (under the same conditions), the savanna-woodland
bistability does not survive at higher rainfall, due to competition
between savanna and forest trees (Figure \ref{fig:Simulation-results-of}B).
Instead, the irregular cycle discussed above appears. Where it is
too dry for forest, savanna tree cover bistability does survive. To
obtain a complete picture of the behavior of the spatial model and
how it differs from the nonspatial model, its bifurcation diagrams
need to be made. A step towards increased realism is then the consideration
of two spatial dimensions instead of one, with a further step towards
increased realism being the verification of how this diagram is affected
by spatial heterogeneity. 

Taking our results reported here together with the simulation results
in our previous work \cite{Wuyts2017} and other recent work \cite{Touboul2018},
the forest-savanna model can produce bimodal tree cover distributions
in a range of external parameters due to: (i) bistability between
savanna and woodland, (ii) existence of forest-savanna cycles, (iii)
spatial heterogeneity of forcings other than rainfall. Fitting our
model for separate regions to data in empirically justified parameter
ranges might reveal differences between different regions or suggest
which model components are not adequately captured. That much of the
tree cover bimodality in the Amazon region can be attributed to spatial
heterogeneity, leaving little remaining bimodality \cite{Wuyts2017},
indicates on one hand that bistability and cyclic behavior play at
most a small role in Amazonia. Nonetheless, dry forests in Amazonia
and elsewhere might still exist as an alternative state to moist forest
and/or savanna. In Africa, where there exist large areas of high tree
cover savannas \cite{Grainger1999} and where fire occurrence is higher
\cite{vanNes2018}, bistability and cyclic behavior can be expected
to play a larger role. A possibility other than the ones hitherto
mentioned is that the observed bimodality is an artifact, resulting
from data algorithms \cite{Hanan2014} or preprocessing methods \cite{Gerard2017}.
Therefore, the multistability hypothesis should be tested on tree
cover data produced with methods that are less likely to generate
such artifacts. 

Finally, there exist other types of feedbacks than assumed here and
which can induce multistability. These include feedbacks between soil
fertility and vegetation \cite{Paiva2015}, rainfall and vegetation
\cite{Eltahir1994,Pielke2001,Poschl2010,Zemp2017}, and, herbivore
presence and vegetation. As all existing feedbacks may interact on
various scales \cite{Scheffer2005}, there is no doubt that tropical
vegetation is not just complex but also complicated. Nevertheless,
the insight from complexity science that complicated dynamics can
emerge from simple rules suggests that they might be less complicated
than we currently think. In the search for such simple rules, spatiotemporal
conceptual models like the one developed here will be indispensable.
On the other hand, even if the rules turn out to be simpler than expected
initially, their resulting dynamics may only be captured realistically
when they are implemented in models that are sufficiently individual
based.

\section*{Acknowledgments}

We thank Jan Sieber for comments and Daniele Avitabile for sharing
and explaining the MATLAB code for numerical continuation. For funding,
BW and ARC acknowledge the UK EPSRC (BW: grant EP/N023544/1). JIH acknowledges the UK NERC (GGR Programme)
and the EU FP7 (project LU4C). 

\newpage{}

{\LARGE{}Supporting Information}{\LARGE\par}

\appendix
\renewcommand{\theequation}{A\arabic{equation}} 
\setcounter{equation}{0}  

\section{Model construction\label{sec:Model-construction}}

\subsection*{General overview }

The model in space and time can be written as a system of stochastic
partial differential equations of the reaction-diffusion type. In
ecological context, one describes the dynamics of a species density
as the sum of a reaction term, representing local demography, and
a diffusion term, representing migration of the species through space
from areas with high to areas with low density. Our model can be most
compactly expressed in vector form\textcolor{red}{{} }
\begin{eqnarray*}
\partial_{t}\mathbf{Y} & = & \mathbf{J}(\mathbf{Y};\mathbf{A})+\mathbf{D}\nabla^{2}\mathbf{Y},
\end{eqnarray*}
where $\mathbf{Y}=(S,T,F,G)$ with $S$ representing savanna sapling
cover density, $T$ savanna tree cover density, $F$ forest tree cover
density, and $G$ grass cover density. $\mathbf{A}$ is a vector of
exogenous environmental variables or parameters that force the system,
such as mean rainfall, rainfall seasonality and soils \cite{Wuyts2017}.
These forcing variables are in general all heterogeneous in space,
but in the main text we kept rainfall seasonality and soils constant,
and took mean annual rainfall as the only heterogeneous forcing. $\mathbf{J}$
is a vector of reaction terms representing local population dynamics
involving both gains and losses and contains nonlinear terms in both
$\mathbf{Y}$ and $\mathbf{A}$. Note that one other forcing variable
is contained in $\mathbf{J}$, being human impact. With regards to
qualitative dynamics and steady state distributions, the particular
choices of the functional forms of $\mathbf{J}$ are arbitrary to
some extent. As long as we choose the right shape, the phase portrait
should be topologically equivalent to the true functional form. $\mathbf{D}$
is a diagonal matrix with diffusion constants. We take the forcing
variables as constant in time. This is done by replacing $\mathbf{A}(\mathbf{x},t)$
by its long-term mean, which is only a function of space. We denote
it further as $\mathbf{A}(\boldsymbol{x})=\mathbf{A}$. We only consider
1D space here. Hence, $\nabla^{2}=\partial_{x}^{2}$, $\mathbf{x}=x$
and $\mathbf{y}=y$. In 2D, front dynamics will be influenced by front
curvature but this is minimal for the spatial scales considered \cite{Wuyts2017c}.

\subsection*{Local rates of change}

Here, we show how the local rates of change $J_{Y}$ for $Y$ any
of the cover types ($S,T,F,G$) are chosen. As in any population model,
we have 
\begin{eqnarray*}
\text{change} & = & \text{gain}-\text{loss}
\end{eqnarray*}
Below, the set of gain processes ${\cal P}_{G}$ contains recruitment
and growth, while that of the loss processes ${\cal P}_{L}$ contains
mortality from competition for resources, drought, fire and human
impact. Each of those processes can be captured with a different term,
such that the equations of the cover types $S,T,F$ and $G$ take
the form 
\[
\dot{Y}=J_{Y}(\mathbf{Y};\mathbf{A})=\sum_{i\in{\cal P}_{G}}G_{Y,i}(\mathbf{Y};\mathbf{A})-\sum_{i\in{\cal P}_{L}}L_{Y,i}(\mathbf{Y};\mathbf{A},z).
\]
As external climatic/edaphic forcing we choose
\[
\mathbf{A}=(MAR,MSI,EFS),
\]
where $MAR$ stands for the observed multi-annual mean of rainfall,
$MSI$ Markham's seasonality index and $EFS$ the edaphic suitability
for forest {[}this lets our model agree with that of \cite{Wuyts2017}
but using a more compact notation{]}. $Y$ refers to any of the cover
types. Functions $G_{Y,i}$ and $L_{Y,i}$ are respectively, total
gains and total losses per time of species $Y$ by process $i$. The
functional forms will be chosen inspired by an understanding of the
effect of all relevant processes. 

Gain functions consist of
\begin{eqnarray*}
\sum_{i\in{\cal P}_{G}}G_{Y,i}(\mathbf{Y};\mathbf{A}) & = & G_{Y,e}(\mathbf{Y};\mathbf{A})+G_{Y,r}(\mathbf{Y};\mathbf{A})+G_{Y,m}(\mathbf{Y};\mathbf{A}),\\
 &  & \text{expansion,\hspace{0.5cm}}\text{recruitment,}{\color{green}\hspace{0.5cm}}\text{\text{mortality other}\text{ types}}
\end{eqnarray*}
Gains can occur due to local (subgrid) expansion of vegetation ($i=e$),
or, in case of an age-structured species, due to recruitment from
a younger stage ($i=r$), or due to increased availability of space
after mortality of other cover types ($i=m$). Loss functions consist
of 
\begin{adjustwidth}{-2.in}{0in}
\begin{eqnarray*}
\sum_{i\in{\cal P}_{L}}L_{Y,i}(\mathbf{Y};\mathbf{A}) & = & L_{Y,r}(\mathbf{Y};\mathbf{A})+L_{Y,c}(\mathbf{Y};\mathbf{A})+L_{Y,d}(\mathbf{Y};\mathbf{A})+L_{Y,f}(\mathbf{Y};\mathbf{A})+L_{Y,h}(\mathbf{Y},z)+L_{Y,o}(\mathbf{Y})\\
 &  & \text{recruitment,\hspace{0.25cm}competition,\hspace{0.25cm}}\text{drought,}\hspace{0.5cm}\text{fire,}\text{\hspace{1.25cm}humans,\hspace{0.75cm}other}
\end{eqnarray*}
\end{adjustwidth}
They can occur due to: recruitment to an older stage ($i=r$), interspecific
competition ($i=c$), drought ($i=d$), human impact ($i=h$) or other
causes ($i=o$). Below, each of the gain/loss terms are discussed.
\begin{itemize}
\item \textit{Cover gains and losses due to expansion.} Gains due to expansion
involve increase of cover area of a more competitive cover type at
cost of the cover area of a less competitive type. Hence, 
\[
G_{Y,e}(\mathbf{Y};\mathbf{A})=R_{Y}(\mathbf{A})Y\mathbf{Y}.\mathbf{v}_{c,Y},
\]
with $Y$ the species that expands its cover and $\mathbf{v}_{c,Y}$
the competitiveness vector for cover type $Y$, which has ones at
its elements where the corresponding cover type is less competitive
than, and can be colonized by, $Y$. The rationale is that colonization
by a species occurs due to an interaction of the space that is occupied
by that species with the space that can be colonized by it. Having
chosen the competitive hierarchy $F>T>S>G$ \cite{Staver2012} in
absence of water limitation, we have 
\begin{eqnarray*}
G_{F,e}(\mathbf{Y};\mathbf{A}) & = & R_{S}(\mathbf{A})T\mathbf{Y}.\mathbf{v}_{c,S}=R_{F}(\mathbf{A})F(G+S+T),\\
G_{S,e}(\mathbf{Y};\mathbf{A}) & = & R_{F}(\mathbf{A})F\mathbf{Y}.\mathbf{v}_{c,F}=R_{S}(\mathbf{A})TG,
\end{eqnarray*}
These are removed from the areas of the less competitive types. Hence,
losses due to expansion of competing types are
\begin{eqnarray*}
L_{T,c}(\mathbf{Y};\mathbf{A}) & = & R_{F}(\mathbf{A})TF\\
L_{S,c}(\mathbf{Y};\mathbf{A}) & = & R_{F}(\mathbf{A})SF\\
L_{G,c}(\mathbf{Y};\mathbf{A}) & = & R_{F}(\mathbf{A})GF+R_{T}(\mathbf{A})GT
\end{eqnarray*}
The only age-structured part of our model is that of savanna saplings
$S$ and savanna (adult) trees $T.$ Note therefore that $G_{S,e}$
has $TG$ instead of $SG$ because expansion only comes from interaction
of adult species $T$ with places that saplings $S$ can colonize
($G$). The dependence of $R_{Y}$ on $\mathbf{A}$ captures the effect
of water availability on growth, which we choose as
\[
R_{Y}(\mathbf{A})=\max[0,r_{Y}(1-e^{-\mathbf{k_{R_{Y}}}.\mathbf{A}+a_{Y}})],
\]
with $r_{Y}$ the maximal growth rate, $\mathbf{k_{R_{Y}}}$ as growth
rate increase for every component of $\mathbf{A}$ and $a_{Y}$ fixing
the rate for $\mathbf{A}=\boldsymbol{0}$. This function captures
saturation of growth rate where water limitation is less severe, which
is supported by empirical work (e.g. saturation of NDVI as a function
of rainfall and high temporal correlations between rainfall and NDVI
below saturation;\textcolor{black}{{} \cite{Wuyts2012,Scheffer2005}}{]}.
\item \textit{Gains and losses due to recruitment}. Saplings will recruit
into adults, such that, 
\[
G_{T,r}(\mathbf{Y};\mathbf{A})=-L_{S,r}(\mathbf{Y};\mathbf{A})=Q(\Phi)S,
\]
where recruitment rate $Q$ is a function of burnt area fraction $\Phi$,
\[
Q(\Phi)=Q_{0}(1-h\Phi).
\]
 $Q_{0}$ is the recruitment rate in absence of fire and $Q_{0}(1-h)$
the recruitment rate in presence of fire for a particular year, with
$0\leq h\leq1$. Hence, in agreement with previous empirical work,
fire affects the establishment rather than the mortality of savanna
trees. 
\item \textit{Base mortality. }This is the base mortality in absence of
fire, drought, competition or human impact. As is common in ecology,
we choose this mortality linear in $Y$, 
\[
L_{Y,o}=m_{Y,o}Y
\]
with $m_{Y,o}$ the base mortality rate of $Y$. 
\item \textit{Losses due to drought}. Drought-related mortality rate $M_{Y}(\mathbf{A})$
will be chosen such that drought effects cause increased mortality
below a certain threshold of \textbf{$\mathbf{A}$}. \textcolor{black}{We
choose} 
\[
L_{Y,d}(Y;\mathbf{A})=M_{Y}(\mathbf{A})Y,
\]
\textcolor{black}{with}
\[
M_{Y}(\mathbf{A})=e^{-\mathbf{k_{M_{Y}}}.\mathbf{A}+a_{M_{Y}}},
\]
or when absorbing the base mortality rate into this term,
\[
M_{Y}(\mathbf{A})=m_{Y,o}+e^{-\mathbf{k}_{M_{Y}}.\mathbf{A}+a_{M_{Y}}}.
\]
Such nonlinear increase of mortality with dryness is assumed to be
a consequence of exceedance of thresholds related to tree water availability.
\item \textit{Losses due to fire}. $\Phi$ is burnt area fraction per year.
Grasses and savanna tree saplings resprout very rapidly so they are
assumed to be unaffected by fire on the considered time scale. Savanna
trees are fire adapted so also they are also assumed to be unaffected
by fire \cite{Staver2012}. Hence, in this model, only forest trees
experience direct mortality due to fire. This mortality is chosen
proportional to burnt area and burning is assumed to occur in a homogeneously
distributed fashion over the pixel. Thus
\[
L_{F,f}(\mathbf{Y};\mathbf{A})=b\Phi F,
\]
where $b$ is the fraction of fire affected forest that dies, also
the fire sensitivity of forest cover. Previous research shows that
fire in the tropics is determined by climate on large scales and by
tree cover on small scales \cite[and the references therein]{Pausas2017}.
On the other hand, local-scale forest distribution affects fire occurrence.
This means that forest and fire interact on local scales. Hence, a
positive feedback can arise if this interaction occurs in a nonlinear
fashion and if it reinforces changes. There exists evidence from independent
lines of research for such positive feedbacks \cite{Pausas2017}.
A fire feedback operating on small spatial scales is crucial for producing
the two stable states, savanna and forest \cite{Staver2012,vanNes2014,Schertzer2015}.
We parameterized the fire feedback by choosing burnt area fraction
$\mathbb{F}$ as a sigmoid-shaped function of fire-prone cover, consistent
with fire percolation models. We let this function further also depend
on climate. The double-striked notation is used to distinguish it
from the functions and variables related to the cover types. The fundamental
process responsible for the fire feedback is small-scale spatial fire
percolation over a fire prone layer. Simulations have shown \cite{Schertzer2015}
that this process induces a sharp increase of fire-related mortality
around the percolation threshold, which occurs when about 60\% of
the landscape is fire-prone. We do not intend to model this percolation
process but take a mesoscale approach, where we choose $\Phi(Y;\mathbf{A})$
to have a sharp increase around a total nonherbaceous cover $1-G-S=T+F=Y_{c}$.
Instead of modeling a positive threshold response of fire on fire-conductive
cover ($S+G$), we choose to formulate the functional form as a negative
threshold response of fire on non-fire-conductive cover ($T+F$).
This makes the analysis easier while keeping the model qualitatively
the same. We hence choose 
\begin{equation}
\Phi(T,F;\mathbf{A})=\frac{1}{\tau}\frac{Y_{c}(\mathbf{A})^{n}}{Y_{c}(\mathbf{A})^{n}+(T+F)^{n}}.\label{eq:BA}
\end{equation}
The exponent $n$ is a positive integer that controls the steepness
of increase of burnt area fraction at $Y\approx Y_{c}$. We chose
$n=4$. $Y_{c}(\mathbf{A})$ captures the varying percolation threshold
with hydrological conditions - a lower threshold in drier environments.
This dependence is chosen to be piecewise linear,
\[
Y_{c}(\mathbf{A})=\text{max}[0,Y_{c,0}+\mathbf{k}_{c}.\mathbf{A}].
\]
Here, $\mathbf{k}_{c}$ is a constant vector and $Y_{c,0}$ a constant
scalar. The elements of $\mathbf{k}_{c}$ represent the sensitivity
of $Y_{c}(\mathbf{A})$ to the different components of $\mathbf{A}$.
$Y_{c}$ has a value of about 40\% for common conditions, which is
the tree cover value at which fire has been observed to increase \cite{Archibald2009a,Staver2011a,Wuyts2017}.
\item \textit{Losses due to human impact}. Deforestation of forest trees
is chosen as
\[
L_{F,h}(F,z)=C(z)F,
\]
where $C(z)$ is the deforestation rate and $z$ distance from anthropogenically
impacted areas. We choose
\[
C(z)=ce^{-k_{C}z},
\]
such that the deforestation rate decays with distance from impacted
areas. $c$ is the the maximum deforestation rate, which occurs in
agricultural areas ($z=0$). 
\item \textit{Gains due to mortality of other cover types}. When any cover
type loses space, it makes place for other cover types. When this
does not occur due to competition or recruitment, grass is the default
cover type that gains ground. This agrees with taking the assumption
that grass grows back instantly, which was also taken in \cite{Staver2012}.
Therefore, 
\[
G_{G,m}(\mathbf{Y};\mathbf{A})=\sum_{\begin{array}{c}
Y\in\{S,T,F\}\\
i\in\{d,f,h,o\}
\end{array}}L_{Y,i}(\mathbf{Y};\mathbf{A}),
\]
of which the terms are defined above. 
\end{itemize}
The system of equations locally obeys the aforementioned mathematical
constraint for every point in time 
\[
\sum_{Y\in\{S,T,F,G\}}Y=1.
\]
Differentiation of the above equation with respect to time yields
\begin{eqnarray}
\sum_{Y\in\{S,T,F,G\}}\partial_{t}Y & = & 0,\nonumber \\
\sum_{Y\in\{S,T,F,G\}}J_{Y}(\mathbf{Y};\mathbf{A}) & = & 0.\label{eq:consJ}
\end{eqnarray}
The conservation equation \ref{eq:consJ} implies that the sum of
all loss and gain terms should be zero. We can see that this is the
case because: (i) total expansion due to successful competition is
at the cost of total loss due to unsuccessful competition, (ii) recruitment
lost by $S$ is gained by $T$, (iii) tree cover losses due to fire,
drought and human impact are gained by grass cover.

\subsection*{Spatial dependence}

Thus far, we have only treated the dynamics as spatially independent
{[}$\mathbf{Y}=\mathbf{Y}(t)${]}, for particular parameter values
of hydrology \textbf{$\mathbf{A}$} and distance to human impacted
areas $z$. To run this model for a whole region {[}$\mathbf{Y}=\mathbf{Y}(\mathbf{x},t)${]},
we need to take into account not only the spatial heterogeneity of
these variables but also the relevant spatial interactions. 
\begin{enumerate}
\item \textit{Spatial heterogeneity.} Climatic, edaphic and anthropogenic
spatial heterogeneity can be included by taking $\mathbf{A}$ and
$z$ as functions of space $\mathbf{A}(\mathbf{x})$ and $z(\mathbf{x})$. 
\item \textit{Spatial interaction.} We assume that diffusion of cover types
only occurs due to spread of seeds. We do not model seed dispersal
but approximate it by dispersal of saplings. Hence the diffusion coefficient
of savanna adult tree cover is zero. That of forest cover is not zero
because part of its population is in the sapling stage. Hence,
\[
\mathbf{D}=(D_{S},0,D_{F},0).
\]
The cover types that diffuse from neighboring areas settle in the
areas that are taken by grasses. Therefore, in the spatial model,
the diffusion terms are also grass cover loss terms due to unsuccessful
competition, or 
\[
L_{G,c}(\mathbf{Y};\mathbf{A})=R_{F}(\mathbf{A})GF+R_{T}(\mathbf{A})GT+D_{S}\nabla^{2}S+D_{F}\nabla^{2}F.
\]
Spatial interaction can also occur due to spread of fire. While analysis
of the model with fire spread is less straight forward \cite{Wuyts2017c},
the conclusions are the same. 
\end{enumerate}

\subsection*{Forest-savanna model}

Here, we develop the forest-savanna model with all previously mentioned
cover types. Note that we do not write the explicit dependence on
space and time. Hence, this means that all cover types are a function
of space $\mathbf{x}$ and time $t$. The forcings are only a function
of space, i.e. $\mathbf{A}=\mathbf{A}(\mathbf{x})$ and $z=z(\mathbf{x})$.
Based on the previous sections, we have
\begin{eqnarray*}
\partial_{t}S & = & G_{S,e}(\mathbf{Y};\mathbf{A})-L_{S,r}(\mathbf{Y};\mathbf{A})-L_{S,d}(S;\mathbf{A})-L_{S,c}(\mathbf{Y};\mathbf{A})+D_{S}\nabla^{2}S,\\
\partial_{t}T & = & G_{T,r}(\mathbf{Y};\mathbf{A})-L_{T,d}(T;\mathbf{A})-L_{T,c}(\mathbf{Y};\mathbf{A}),\\
\partial_{t}F & = & G_{F,e}(\mathbf{Y};\mathbf{A})-L_{F,d}(F;\mathbf{A})-L_{F,f}(\mathbf{Y};\mathbf{A})-L_{F,h}(F,z)+D_{F}\nabla^{2}F,\\
\partial_{t}G & = & -L_{G,c}(\mathbf{Y};\mathbf{A})+G_{G,m}(\mathbf{Y};\mathbf{A}),
\end{eqnarray*}
Filling in the gains and losses and making use of $G=1-S-T-F$, we
obtain 

\begin{eqnarray*}
\partial_{t}S & = & R_{s}(\mathbf{A})(1-S-T-F)T-Q[\Phi(T,F;\mathbf{A})]S-M_{S}(\mathbf{A})S-R_{F}(\mathbf{A})SF,\\
\partial_{t}T & = & Q(\Phi)S-M_{T}(\mathbf{A})T-R_{F}(\mathbf{A})TF,\\
\partial_{t}F & = & R_{F}(\mathbf{A})(1-F)F-b\Phi(T,F;\mathbf{A})F-M_{F}(\mathbf{A})F-C(z)F+D_{F}\nabla^{2}F,
\end{eqnarray*}
We briefly remind the reader of some of the model parameters shown
here. This model is forced by the spatial distribution of $\mathbf{A}$
and the distance to human impact $z$. $b$ is the (constant) sensitivity
of forest cover to fire. $Q$ represents sapling recruitment into
adults and is a linearly decreasing function of burnt area fraction.
$C(z)$ is the deforestation rate which decays with $z$. Note that
the simulation model used to produce Figure \ref{fig:Front-between-forest}A
has no deforestation term in the equation of $\partial_{t}T$. When
hiding the dependence on $\mathbf{A}$, we have
\begin{eqnarray}
\partial_{t}S & = & R_{s}(1-S-T-F)T-Q[\Phi(T,F)]S-M_{S}S-R_{F}SF+D_{S}\nabla^{2}S,\nonumber \\
\partial_{t}T & = & Q[\Phi(T,F)]S-M_{T}T-R_{F}TF,\nonumber \\
\partial_{t}F & = & R_{F}(1-F)F-b\Phi(T,F)F-M_{F}F-C(z)F+D_{F}\nabla^{2}F,\label{eq:fullnse}
\end{eqnarray}

\section{Forest growth rate $r_{F}$\label{sec:Forest-growth-rate}}

Here, we will show how we derived the maximum forest growth rate.
We do this to set the time scale of the model dynamics (other parameters
were initially estimated relative to $r_{F}$). The steady state forest
cover value under sufficiently moist conditions is about 80\%. Therefore,
we made sure that this also occurs in the model by first seeing that
in moist conditions far from human-impacted areas, $C_{F}=0$, $\Phi=0$,
$R_{F}=r_{F}$ and $M_{F}=m_{F,o}$ such that 
\begin{eqnarray*}
\frac{dF}{dt} & = & r_{F}(1-F)F-m_{F,o}F.
\end{eqnarray*}
We will further set $m_{F,o}\equiv m$ and $r_{F}\equiv r$. The ODE
can be solved by separation of variables using partial fractions such
that the time that forest needs to grow from $F_{0}$ to $F_{1}>F_{0}$
is 
\begin{equation}
t_{1}-t_{0}=\frac{1}{r(1-m)}\text{log}(\frac{\left|1-m-F_{0}\right|F_{1}}{\left|1-m-F_{1}\right|F_{0}}).\label{eq:dtF}
\end{equation}
If we take as initial tree cover a small value that could result from
noise and as final tree cover the carrying capacity, we have 
\begin{eqnarray*}
F_{0} & = & 0.01,\;F_{1}=0.8.
\end{eqnarray*}
At carrying capacity, we have $r(1-F)F-mF=0,$such that 
\[
F^{*}=1-\frac{m}{r},\;F^{*}=0.
\]
of which only the first is stable. As the data shows that $F^{*}$
has to be equal to $0.8$, such that $m=0.2r$. Substituting this
and the values for $F_{0},F_{1}$, we obtain 
\begin{equation}
t_{1}-t_{0}=\frac{1}{r(1-0.2r)}\text{log}(400\frac{0.99-0.2r}{1-r}).\label{eq:r_dt}
\end{equation}
This function is plotted in Figure \ref{fig:Recovery-time-of}. In
a recent study on recovery of secondary forests \cite{Poorter2016},
it was found that moist forests regain the median value of old-growth
forests after about at least 30 years. Therefore, we chose the $r$
value consistent with this time, which is 
\[
r\approx0.2.
\]

\newpage{}

\renewcommand{\thetable}{A\arabic{table}} 
\renewcommand{\thefigure}{A\arabic{figure}} 
\setcounter{figure}{0} 
\setcounter{table}{0}

\section*{Supplementary Table}
\noindent \begin{flushleft}
\begin{table}[!ht]
\begin{adjustwidth}{-1.0in}{0in}
\noindent \begin{centering}
\begin{tabular}{cccc}
\hline 
process and equation & value & parameter & units\tabularnewline
\hline 
cover expansion rate & 0.09,0.20 & $r_{S},r_{F}$ & $\text{y}^{-1}$\tabularnewline
$R_{Y}(\mathbf{A})=\max[0,r_{Y}(1-e^{-\mathbf{k_{R_{Y}}}.\mathbf{A}+a_{R_{Y}}})]$ & (0.005,-,-) & $\mathbf{k}_{R_{S}}$ & $(\text{mm\ensuremath{^{-1}}},\text{-,-})$\tabularnewline
 & (0.003,3.26,-) & $\mathbf{k}_{R_{F}}$ & $(\text{mm\ensuremath{^{-1}}},\text{-,-})$\tabularnewline
 & 0.25,0.196 & $a_{R_{S}},a_{R_{F}}$ & -\tabularnewline
cover reduction rate by drought & 0.023,0.041 & $m_{S,o}=m_{T,o},m_{F,o}$ & $y^{-1}$\tabularnewline
$M_{Y}(\mathbf{A})=m_{Y,o}+e^{-\mathbf{k}_{M_{Y}}.\mathbf{A}+a_{M_{Y}}}$ & -,-2.15 & $a_{M_{S}}=a_{M_{T}},a_{M_{F}}$ & -\tabularnewline
 & (0.008,-,-) & $\mathbf{k}_{M_{S}}=\mathbf{k}_{M_{T}}$ & $(\text{mm\ensuremath{^{-1}}},\text{-,-})$\tabularnewline
 & (0.008,-4.66,1.5) & $\mathbf{k}_{M_{F}}$ & $(\text{mm\ensuremath{^{-1}}},\text{-,-})$\tabularnewline
\begin{tabular}{c}
savanna tree cover recruitment rate\tabularnewline
$Q(\Phi)=Q_{0}(1-h\Phi)$\tabularnewline
\end{tabular} & 0.04,0.85 & $Q_{0},h$ & $\text{y}^{-1},\text{-}$\tabularnewline
\begin{tabular}{c}
local burnt area fraction\tabularnewline
$\Phi(T,F;\mathbf{A})=\frac{1}{\tau}\frac{Y_{c}(\mathbf{A})^{n}}{Y_{c}(\mathbf{A})^{n}+(T+F)^{n}},$\tabularnewline
\end{tabular} & 2.7,4 & $\tau,n$ & $\text{y}^{-1},\text{-}$\tabularnewline
with: $Y_{c}(\mathbf{A})=\text{max}[0,Y_{c,0}+\mathbf{k}_{c}.\mathbf{A}]$ & 0.484 & $Y_{c,0}$ & -\tabularnewline
 & (-1.43e-04,.2,-.1) & $\mathbf{k}_{c}$ & $(\text{mm\ensuremath{^{-1}}},\text{-,-})$\tabularnewline
forest cover fire sensitivity & 0.46 & $b$ & -\tabularnewline
\begin{tabular}{c}
deforestation rate\tabularnewline
$C(z)=ce^{-k_{C}z}$\tabularnewline
\end{tabular} & 0.092,0.0015 & $c,k_{C}$ & -,$\text{m}^{-1}$\tabularnewline
diffusion coefficient of $F,S$ & 0.1,0.2 & $D_{F},D_{S}$ & $\text{km}^{2}\text{y}^{-1}$\tabularnewline
\hline 
\end{tabular}
\par\end{centering}
\caption[Model parameters]{{\small{}Model parameters of the forest-savanna model {[}equation
(\ref{eq:fullnse}){]}. $\mathbf{A}=\mathbf{A}(\mathbf{x})$. The
components of are: $A_{1}=P$ (mean annual rainfall), $A_{2}=M$ (Markham's
seasonality index), $A_{3}=\pi-\bar{\pi}$ (edaphic forest suitability).
$\pi$ captures the effect of soils on forest occurrence and is taken
from \cite{Wuyts2017}, i.e. $A_{3}=0.00238\varphi_{s}-0.188\varphi_{c}-5.99\rho-0.183\varphi_{c}\rho+6.39$,
where $\rho$ is topsoil bulk density, $\varphi_{s}$ topsoil sand
fraction, and $\varphi_{c}$ topsoil clay fraction. The components
of the vectors $\boldsymbol{k}_{i}$ multiply the components of $\mathbf{A}$.
If a component is indicated as '-', the considered equation is not
a function of the corresponding component of $\mathbf{A}$.}}
\label{tab:Model-parameters-of}
\end{adjustwidth}
\end{table}
\par\end{flushleft}

\newpage{}
\section*{Supplementary Figures}

\begin{figure}[!h]
\begin{adjustwidth}{-1.0in}{0in}
\begin{centering}
\includegraphics[width=.8\columnwidth]{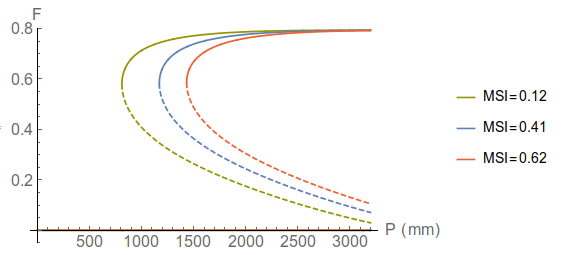}
\par\end{centering}
\caption{Homogeneous steady states (HSS) of forest cover ($F$) in the forest
model without human impact {[}$C(z)=0$ in (\ref{eq:FG}){]} as a
function of mean annual rainfall ($P$) for average soils and with
rainfall seasonality ($MSI$) as indicated in the legend. Stable states
are indicated with solid lines and unstable steady states with dashed
lines. HSS are steady states of the nonspatial model ($\delta=D_{F}=0$).
These plots were obtained by finding the roots of the reaction term
in (\ref{eq:FG}). The stable branches (solid) are metastable states
in the spatial model - they can persist if the whole domain is in
the same state and if they are not exposed to perturbations larger
than a small threshold. }
\label{fig:Homogeneous-steady-states}
\end{adjustwidth}
\end{figure}

\begin{figure}[!h]
\begin{adjustwidth}{-1.0in}{0in}
\begin{centering}
\includegraphics[width=.65\columnwidth]{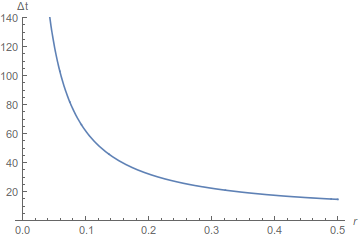}
\par\end{centering}
\caption{Recovery time of undisturbed moist forest as a function of parameter
$r_{F}$ when taking $F^{*}=0.8$ and when the initial forest cover
$F_{0}=0.01$, based on (\ref{eq:r_dt}).}
\label{fig:Recovery-time-of}
\end{adjustwidth}
\end{figure}

\newpage{}

\renewcommand\refname{Literature Cited}
\bibliographystyle{plos2015}

\end{document}